\documentclass[reprint,
superscriptaddress,
 amsmath,amssymb,
 aps,
prb,
]{revtex4-1}

\usepackage{comment}
\usepackage[inline, shortlabels]{enumitem}
\usepackage{graphicx}% Include figure files
\usepackage{dcolumn}% Align table columns on decimal point
\usepackage{bm}% bold math
\usepackage{xcolor}

\begin{document}
% Average Cluster Method

\preprint{APS/123-QED}

\title{An averaged cluster approach to including chemical short range order in KKR-CPA}

\author{Vishnu Raghuraman}
\affiliation{Department of Physics, Carnegie Mellon University, Pittsburgh, PA, 15213, USA}
\author{Yang Wang}
\affiliation{
Pittsburgh Supercomputing Center, Carnegie Mellon University, Pittsburgh, PA, 15213, USA
}%
\author{Michael Widom}
\affiliation{Department of Physics, Carnegie Mellon University, Pittsburgh, PA, 15213, USA}
\begin{abstract}
    The single-site Korringa-Kohn-Rostoker Coherent Potential Approximation (KKR-CPA) ignores short range ordering present in disordered metallic systems. In this paper, we establish a new technique to fix this shortcoming by embedding an averaged cluster that displays chemical short range order (SRO). The degree of SRO can be tuned by externally defined order parameters. This averaged cluster can be embedded in the single site CPA medium, or a self-consistently obtained effective medium that contains SRO information. The validity of this method is demonstrated by applying it to two alloy systems - the CuZn body centered cubic (BCC) solid solution, and AlCrTiV, a four-element BCC high entropy alloy. A comparison between the non-self-consistent and self-consistent modes is also provided for the two above mentioned systems. We make the code available on the internet. Planned extensions to this work are discussed.
\end{abstract}

\maketitle
\section{Introduction}
The KKR-CPA method \cite{kkr-cpa-1,kkr-cpa-2,kkr-cpa-3} is heavily used to study the electronic structure of disordered systems. It is based on obtaining an effective medium, calculated using a single site approximation that mimics the ensemble average of the different possible configurations. CPA successfully predicts the total energy, the density of states, and other system properties of random alloys \cite{kkr-cpa-dos-1,kkr-cpa-dos-2,kkr-cpa-dos-3,kkr-cpa-dos-4}. One major shortcoming of the single site approximation is that it is unable to reproduce chemical ordering that may be present in the system. Short range ordering (SRO) can significantly impact the chemical and mechanical properties of a solid. For example, in the Ti-6Al alloy, experiment confirms that SRO increases the nominal yield stress \cite{sro-ti6al}. In the CoCrNi medium entropy alloy, SRO increases the yield strength by approximately 25\%, increases the nanoindentation hardness, and significantly affects the onset of plasticity \cite{sro-cocrni}. In $\alpha$-brass, SRO can cause a change in the resistivity \cite{sro-brass-resistivity} and increase the stacking fault energy \cite{sro-brass-sfe}

We wish to study the effects of SRO within the CPA framework. Modifications to the conventional CPA method are required in order to model chemical ordering. Early efforts include the molecular CPA (MCPA) \cite{mcpa} method, where the crystal is divided into cells and each cell is treated like a ``molecule''. The vibrational spectra of a one-dimensional isotopically disordered chain was successfully calculated using MCPA. For more complex systems, this method is difficult to implement. Gonis \textit{et al} developed the Embedded Cluster Method (ECM-CPA) \cite{pre-ecm-cpa,ecm-cpa} which embeds differently configured clusters of real atoms within a single-site CPA medium. The site-diagonal element of the Green's Function (corresponding to the central atom of the cluster) is used for calculating system properties. This method was used to calculate the DOS for Ag$_c$Pd$_{1-c}$ binary by embedding multiple 13-atom clusters with various configurations (central atom plus 12 nearest neighbors). While this is an extremely powerful approach, it does not scale well with the number of configurations. 

The Locally Self-Consistent Green's Function \cite{lsgf,lsgf-2} (LSGF) method also accounts for local behaviour. This is a supercell approach where the Green's Function for each atom is calculated using a cluster consisting of the atom and it's neighbors, embedded in an effective medium. This cluster is referred to as the local interaction zone (LIZ), which is a crucial concept, originally introduced by the Locally Self-consistent Multiple Scattering (LSMS) method \cite{LSMS}, in the the Green function based, linear scaling supercell approach to the {\it ab initio} electronic structure calculations. Even though the LSGF calculation can be done efficiently, a supercell is more computationally demanding than a CPA calculation where a single atom unit cell is sufficient.

More recent methods include the Non-Local Coherent Potential Approximation (NLCPA) \cite{nlcpa}, which is based on the Dynamical Cluster Approximation \cite{dca} and involves finding a new effective medium, effective structure constant and a cluster-renormalized interactor that includes non-local effects. This is achieved by a self-consistent process where, for a fixed cluster size, all possible clusters are embedded and the impurity path operator is averaged to obtain the path operator for the NL-CPA medium. Unlike the Embedded Cluster approach, NL-CPA is fully self-consistent. This approach was used successfully to obtain the DOS of the CuZn binary using two atom clusters \cite{nlcpa-2,msnlcpa}. Marmodoro \textit{et al} \cite{msnlcpa} improved NL-CPA, making it applicable to multiple sublattices and complex geometries (MSNLCPA). Further improvements to the approach were proposed and applied to iron-based superconductor FeSe \cite{nlcpa-improved}. While NL-CPA has several desirable features, it is difficult to apply to systems that have a very large configuration space.

In the present work, we introduce the Cluster Averaged Coherent Potential Approximation (CA-CPA), that embeds a single ``averaged'' cluster containing SRO within the single site CPA medium. SRO is modelled by a set of pre-determined order parameters. Like ECM-CPA, our theory uses the diagonal element of the Green's Function corresponding to the central atom of the cluster to calculate system properties. This approach does not require an ensemble of specific cluster configurations. It can easily be extended to high entropy alloy \cite{hea-1,hea-2} systems consisting of four or more chemical species. Unlike ECM-CPA, our method can be made self-consistent. By embedding the averaged cluster repeatedly, and solving for the effective medium t-matrix, a self-consistent CPA medium can be calculated. This new effective medium can be used instead of the single site CPA medium. 

The paper is organized as follows. First we provide a brief introduction to the KKR method and the equations needed to embed a cluster in the CPA medium. We then introduce the formalism behind CA-CPA and compare it to NLCPA. We derive an iteration scheme that can be used to obtain an SRO effective medium, and the effect of self-consistency is observed by comparing with the non-self-consistent embedding of the averaged cluster in the single-site CPA medium. We apply our method to calculate density of states and total energy of the CuZn binary alloy and the AlCrTiV high entropy alloy. Both systems show B2 (CsCl) type ordering at low temperatures and A2 (BCC) type ordering at higher temperatures. We compare our results with single site CPA calculations to observe the effects of SRO.  We will also demonstrate that there is little difference between the self-consistent and non-self-consistent cluster embedding. Finally we conclude with planned extensions to this work.

\section{Formalism}
\begin{comment}
\textcolor{red}{Give a brief discussion of DFT approach to condensed matter and show the Kohn-Sham equation with one-electron effective potential $V_{\rm eff}$...} 
\end{comment}
The first principles electronic density functional theory (DFT) approach is based on solving a single electron Schr\"{o}dinger equation, called the Kohn-Sham equation \cite{dft-1, dft-2} (with $\hbar = 1$ and $m_e = \frac{1}{2}$),
\begin{equation}
    \left[-\nabla^{2} + V_{\rm{eff}}(\left[\rho(\bm{r})\right])\right]\psi_{i}(\bm{r}) = \epsilon_{i}\psi_{i}(\bm{r}).
\end{equation}
The Hamiltonian is a functional of the density, given by
\begin{equation}
    \rho(\bm{r}) = \sum_{\substack{ i \\ \epsilon_i\leq\epsilon_{\rm F} }} \left|\psi_i(\bm{r})\right|^{2},
\end{equation}
provided the Kohn-Sham orbital wave functions are orthonormal and the Fermi Energy $\epsilon_{F}$ is determined by the number of electrons in the system. The effective potential $V_{\rm{eff}}$ includes the Hartree potential, which describes the electronic Coulomb repulsion, the electron-nucleus interaction, and the exchange-correlation (XC) potential, a functional of the electron density. With the exception of approximations applied to the XC functional, DFT is an exact theory. In the computational treatment of the XC functional, the  local density approximation (LDA) \cite{dft-1} assumes that the XC functional is only dependent on the local electron density, while the Generalized Gradient Approximation (GGA) \cite{gga1,gga2} assumes dependence on both local density and it's gradient. In our work we choose to apply the GGA.

Solving the Kohn-Sham equation can be done in multiple ways. Because it is an eigenvalue equation, diagonalizing the Hamiltonian is a popular method of obtaining the required energy and wavefunctions. However, an alternate technique, called the KKR-Green's Function Method \cite{faulkner_stocks_wang} can also be used.

\subsection{KKR-Green's Function Method} 

In the KKR method \cite{korringa_1947, kohn_rostoker_1954}, the system is divided into cells, each of which is centered around an atom. The one-electron effective potential $V_{\rm eff}$ is a sum of localized potentials, $v_n$, within each cell. Specifically, for a cell $n$ with volume $\Omega_{n}$ whose center is the atomic site determined by position vector $\bm{R_n}$, the local potential is given by
\begin{equation}
    v_n(\bm{r_n}) = 
    \begin{cases}
    V_{\rm{eff}}(\bm{r}), & \text{if}\;\bm{r} \in \Omega_n \\
    0, & \text{otherwise}
    \end{cases},
\end{equation}
where $\bm{r_n} = \bm{r} - \bm{R_n}$. Consider each cell as an electron scattering center. The multiple scattering path matrix $\underline{\tau}^{nm}$ \cite{tau_def}, defined as the sum of all scattering processes that start from cell $n$ and end at cell $m$, becomes
\begin{equation}
    \underline{\tau}^{nm}(\epsilon) = \underline{t}^{n}(\epsilon)\delta_{nm} + \underline{t}^{n}(\epsilon)\sum_{k\neq n} \underline{g}^{nk}(\epsilon) \underline{\tau}^{km}(\epsilon),
    \label{eq:tautrecur}
\end{equation}
where $\underline{t}^{n}(\epsilon)$ represents the single site scattering t-matrix associated with potential $v_n$ and $\underline{g}^{nk}(\epsilon)$ is the free electron propagator matrix that describes the propagation of a free electron with energy $\epsilon$ from site $n$ to site $k$. 

Calculating the multiple scattering path matrix is important because the Green's function in cell $n$, given by $G(\bm{r_n}, \bm{r_n}, \epsilon)$ can be written as \cite{faulkner_stocks_1980,faulkner_stocks_wang} 
\begin{eqnarray}
    G(\bm{r_n}, \bm{r_n}, \epsilon) & = & \sum_{LL^{\prime}} Z^{n}_L (\bm{r_n}, \epsilon)\tau^{nn}_{LL^{\prime}}(\epsilon)Z^{n\bullet}_{L^{\prime}} (\bm{r_n}, \epsilon) \nonumber \\
& &   - \sum_{L} Z^{n}_L (\bm{r_n}, \epsilon)J^{n\bullet}_L(\bm{r_n}, \epsilon),\label{eq:gf}
\end{eqnarray}
where $L$ is a combination of orbital angular momentum quantum number $l$ and magnetic quantum number $m$. $Z^{n}_L(\bm{r_n}, \epsilon)$ and $J^{n}_L(\bm{r_n}, \epsilon)$ represent the regular and irregular local solutions to Schr\"{o}dinger's equation in cell $n$ with proper boundary conditions near $r_n=0$ and at the bounding sphere of cell $n$, respectively. The bullet symbol ``$\bullet$" in Eq.~(\ref{eq:gf}) represents complex conjugate applied to the spherical harmonics contained in $Z^{n}_L(\bm{r_n}, \epsilon)$ and $J^{n}_L(\bm{r_n}, \epsilon)$. The electron density associated with the valence states in cell $n$ can be calculated from the Green's Function by taking the imaginary part of the trace integrated in the valence energy band, from the bottom of the band $\epsilon_{B}$ to the Fermi energy $\epsilon_{F}$,
\begin{equation}
    \rho^{n}(\bm{r_n}) = -\frac{1}{\pi}\mathrm{Im~Tr}\int_{\epsilon_{B}}^{\epsilon_{F}} G(\bm{r_n}, \bm{r_n}, \epsilon)\;d\epsilon.
\end{equation}
This means that calculation of the Kohn-Sham orbital wave functions is unnecessary in the KKR-Green's Function Method. This is advantageous, as  time-consuming operations like orthogonalizing and normalizing the wave functions can be avoided.

\subsection{Cluster Impurity Equations}
In certain cases, the calculation of the multiple scattering path matrix (Eq.~\ref{eq:tautrecur}) can be simplified. For a single site CPA medium, all the cells have the same t-matrix, given by $\underline{t}_{\rm{CPA}}(\epsilon)$. In such a system, the multiple scattering path matrix $\underline{\tau}_{\rm CPA}(\epsilon)$ at a particular site $n$ can be written as
\begin{equation}
    \underline{\tau}_{\rm{CPA}}^{nn}(\epsilon) = \frac{1}{\Omega_{BZ}}\int_{\Omega_{BZ}} d^{3}\bm{k} \left[\underline{t}_{\rm{CPA}}^{-1}(\epsilon) - \underline{g}(\bm{k}, \epsilon)\right]^{-1},
\end{equation}
where $\underline{t}_{\rm{CPA}}(\epsilon)$ represents the single-scattering t-matrix for the CPA medium and $\underline{g}(\bm{k}, \epsilon)$ is the lattice Fourier transform of free electron propagator $\underline{g}^{nk}(\epsilon)$. 

Another special case is a crystal with substitutional impurities. Consider a crystal made of the CPA medium at every site except for an impurity at site $n$. If the t-matrix at the impurity site is given by $\underline{t}_{I}(\epsilon)$ , it can be shown that the multiple scattering path matrix $\underline{\tau}_{I}^{nn}(\epsilon)$ at the impurity site is given by
\begin{equation}
    \underline{\tau}_{I}^{nn}(\epsilon) = \left[1 + \underline{\tau}^{nn}_{\rm{CPA}}(\epsilon)(\underline{t}_{I}^{-1}(\epsilon) - \underline{t}_{\rm{CPA}}^{-1}(\epsilon))\right]^{-1}\underline{\tau}_{\rm{CPA}}^{nn}(\epsilon).
    \label{eq:singlesite}
\end{equation}
This idea can be extended to a cluster impurity in the CPA medium. If we embed a cluster of size $N_c$, $\underline{\tau}_{I}^{nn}(\epsilon)$ at impurity site $n$ is given by
\begin{equation}
    \underline{{\tau}}_{I}^{nn}(\epsilon) = \left(\left[1 + \underline{\underline{\tau}}_{\rm{CPA}}(\epsilon)(\underline{\underline{T}}^{-1}_{I}(\epsilon) - \underline{\underline{T}}^{-1}_{\rm{CPA}}(\epsilon))\right]^{-1}\underline{\underline{\tau}}_{\rm{CPA}}(\epsilon)\right)^{nn}.
    \label{eq:clusterimpurityeqn}
\end{equation}
Here the double underlined terms represent block matrices of size $N_c \times N_c$. The $(n, m)$ block of $\underline{\underline{\tau}}_{\rm{CPA}}(\epsilon)$ is given by \cite{ecm-cpa}
\begin{equation}
    \underline{\tau}_{\rm{CPA}}^{nm}(\epsilon) = \frac{1}{\Omega_{BZ}}\int d^{3}\bm{k} \left[\underline{t}_{\rm{CPA}}^{-1}(\epsilon) - \underline{g}(\bm{k}, \epsilon)\right]^{-1} e^{i\bm{k}\cdot(\bm{R_n} - \bm{R_m})},
    \label{eq:blockTauCPA}
\end{equation}
and $\underline{\underline{T}}^{-1}_{I}(\epsilon)$ and $\underline{\underline{T}}^{-1}_{\rm{CPA}}(\epsilon)$ are block diagonal matrices defined as 
\begin{align}
    \left[\underline{\underline{T}}^{-1}_{I}(\epsilon)\right]^{nm} &= \underline{t}_{I_n}^{-1}(\epsilon)\delta_{nm} \\
    \left[\underline{\underline{T}}^{-1}_{\rm{CPA}}(\epsilon)\right]^{nm} &= \underline{t}^{-1}_{\rm{CPA}}(\epsilon)\delta_{nm}.
    \label{eq:TCPA}
\end{align}
$I_n$ refers to the impurity present at site $n$, and $\underline{t}_{I_{n}}(\epsilon)$ is the corresponding single scattering t-matrix.

\subsection{Averaged Cluster Embedding}
\begin{comment}
\textcolor{red}{Need some good sentences to make a smooth transition, e.g., by starting with a brief discussion of theoretical approaches to including SRO.}
\end{comment}
 The ECM-CPA method discussed in the introduction uses the equations defined in the above subsection to embed a cluster of real atoms in single site CPA medium. Using this approach to study SRO would require embedding many real atom clusters, drawn from an ensemble of configurations weighted by the desired SRO. For large cluster sizes, and for systems that have more than two chemical species, this approach is highly inconvenient. Our idea is to embed clusters consisting of central atoms surrounded by "averaged" neighbor atoms. These neighbors atoms are an average of the chemical species present in the system, weighted by the chemical short-ranged order with single scattering t-matrix
\begin{equation}
    \bar{\underline{t}}_{a}(\epsilon) = \sum_b w_{ab}\; \underline{t}_{b}(\epsilon)
\end{equation}
for $a,\;b$ in the set of chemical species present. The SRO weights $w_{ab}$ are the fraction of neighbors of species $a$ that are species $b$. For a system that shows strong short range ordering, \textit{i.e} unlike neighbors are preferred, $w_{ab}$ approaches $1 - \delta_{ab}$ (taking $\delta_{ab}$ small). For the opposite case of short range clustering, where like neighbors are preferred, $w_{ab}$ approaches $\delta_{ab}$. In the limit of complete disorder, $w_{ab} = \frac{1}{N_{s}}$, where $N_s$ is the number of chemical species present. 

For an equiatomic binary system with species $a$ and $b$, we can transform $w_{ab}$ to the well known Warren-Cowley SRO parameter \cite{warren-cowley}
\begin{equation}
    \alpha_{ab} = 1 - 2w_{ab}.
\end{equation}
For $w_{ab}> 1/2$, $\alpha_{ab}$ becomes negative and approaches -1 as $w_{ab}$ approaches 1 corresponding to short range ordering. For $w_{ab}<1/2$, $\alpha_{ab}$ is positive and approaches 1 as $w_{ab}$ approaches 0 corresponding to short range clustering. At $w_{ab} = 1/2$, $\alpha = 0$ corresponding to complete disorder.

To obtain the multiple scattering path matrix for an average cluster embedded in the CPA medium, we use the cluster impurity equations defined in Section II.B. Consider a lattice for which a particular atom (the central atom) of species $a$ has $\gamma$ nearest neighbors. Then  for $n, m \in \{1,2,\cdots,\gamma\}$ we define the block-matrix $\underline{\underline{T}}_{a}(\epsilon)$ as 
\begin{equation}
   \left[\underline{\underline T}_{a}(\epsilon)\right]^{nm} = 
   \begin{cases}
   \underline{t}_{a}(\epsilon)\delta_{nm}, & \text{if}\;n = 1; \\
   \bar{\underline{t}}_{a}(\epsilon)\delta_{nm},&\text{otherwise.}
   \end{cases}
\end{equation}
Here $n = 1$ is the central atom and the remaining values of $n$ denote the neighbor atoms. Using the definition of $\underline{\underline{T}}_{\rm{CPA}}(\epsilon)$ from Eq.~(\ref{eq:TCPA}) and the block $\underline{\underline{\tau}}_{\rm{CPA}}(\epsilon)$ from Eq.~(\ref{eq:blockTauCPA}), we write the block $\underline{\underline{\tau}}_{a}(\epsilon)$ for our averaged cluster embedded medium as
\begin{equation}
    \underline{\underline{\tau}}_{a}(\epsilon) = \left[1 + \underline{\underline{\tau}}_{\rm{CPA}}(\epsilon)(\underline{\underline{T}}_{a}^{-1}(\epsilon) - \underline{\underline{T}}_{\rm{CPA}}^{-1}(\epsilon))\right]^{-1}\underline{\underline{\tau}}_{\rm{CPA}}(\epsilon).
    \label{eq:taualpha}
\end{equation}
The (11) block $[\underline{\underline{\tau}}_{\alpha}(\epsilon)]^{11}$ corresponding to the central atom is used to calculate the Green's function (\ref{eq:gf}), similar to the ECM-CPA approach. This method can be thought of as a modified ECM-CPA, where instead of embedding a cluster consisting of real atoms, a carefully constructed averaged cluster is embedded. The computationally intensive steps in CA-CPA are (\ref{eq:blockTauCPA}), which is the Brillouin zone integration, and (\ref{eq:taualpha}), which is an O$(N^3)$ operation as it involves the inverse of a large block matrix.

\subsection{Extension to Next Nearest Neighbors}
\label{sec:NNN}
In the above description the cluster size is limited to nearest neighbors, but the method can easily be extended to include longer ranged ordering. To include next nearest neighbors, a new type of ``averaged'' atom has to be defined using a new set of SRO parameters. For a central atom $a$, the next nearest neighbor average atom $\tilde{t}_{a}(\epsilon)$ can be defined as
\begin{equation}
    \tilde{\underline{t}}_{a}(\epsilon) =  \sum_{\beta}w^{\prime}_{ab}\underline{t}_{b}(\epsilon)
\end{equation}
Here $w_{ab}^{\prime}$ represent the next nearest neighbor SRO parameters. The block $\underline{\underline{T}}_{a}(\epsilon)$ is now defined as
\begin{equation}
    \left[\underline{\underline{T}}_{\alpha}(\epsilon)\right]^{nm} = 
    \begin{cases}
    \underline{t}_{a}(\epsilon) \delta_{nm}, & \text{if}\;n = 1 \\
    \bar{\underline{t}}_{a}(\epsilon)\delta_{nm}, & n \in \text{nearest neighbors} \\
    \tilde{\underline{t}}_{a}(\epsilon)\delta_{nm}, & n \in \text{next nearest neighbors}
    \end{cases}
\end{equation}
The $\underline{\underline{\tau}}_{a}(\epsilon)$ matrix can then be obtained from (\ref{eq:taualpha})

If the neighbor atoms are chosen to be single site CPA \textit{i.e}, if the block $\underline{\underline{T}}_{a}(\epsilon)$ is written as
\begin{equation}
   \left[\underline{\underline{T}}_{a}(\epsilon)\right]^{nm} = 
   \begin{cases}
   \underline{t}_{a}(\epsilon)\delta_{nm}, & \text{if}\;n = 1 \\
   \underline{t}_{\rm{CPA}}(\epsilon)\delta_{nm},&\text{otherwise}
   \end{cases},
   \label{eq:limitingcase}
\end{equation}
it is easy to verify that equation (\ref{eq:taualpha}) which represents the cluster $\underline{\underline{\tau}}_{a}(\epsilon)$ reduces to the single site $\underline{\tau}_{a} (\epsilon)$, represented by equation (\ref{eq:singlesite}). This is expected, because embedding a cluster with neighbor atoms as single site CPA is equivalent to embedding a single atom in the CPA medium. An extension to include multipoint correlation functions is conceivable.
\subsection{Self-consistency}
It is possible to extend the CA-CPA method self-consistently. The single-site approximation for the CPA can be expressed in terms of multiple scattering matrices as 
\begin{equation}
    \underline{\tau}_{\rm{CPA}}(\epsilon) = \sum_{a} c_{a} \underline{\tau}_{a} (\epsilon),
    \label{eq:singlesiteapprox}
\end{equation}
where $\underline{\tau}_a (\epsilon)$ is calculated by embedding an atom of species $a$ in the medium and employing (\ref{eq:singlesite}). If an averaged cluster with central atom $a$ at site $n$ is embedded (instead of a single atom), equation (\ref{eq:singlesiteapprox}) can be simply modified to
\begin{equation}
    \underline{\tau}_{\rm{CPA}}^{nn} (\epsilon) = \sum_{a} c_a \underline{\tau}_{a}^{nn}(\epsilon),
    \label{eq:clusterapprox}
\end{equation}
where $\underline{\tau}_{a}^{nn}(\epsilon)$ is now calculated from (\ref{eq:clusterimpurityeqn}).  Averaging as in (\ref{eq:clusterapprox}), we obtain an iteration scheme for $\underline{\tau}_{a}^{nn}(\epsilon)$. Inserting (\ref{eq:clusterimpurityeqn}) in (\ref{eq:clusterapprox}) and defining $\underline{\underline{\Delta}}_{a}(\epsilon) = \underline{\underline{T}}_{a}^{-1}(\epsilon) - \underline{\underline{T}}_{\rm{CPA}}^{-1}(\epsilon)$ yields
\begin{equation}
    \underline{\tau}_{\rm{CPA}}^{nn}(\epsilon) = \sum_{a} c_a \left(\left[1 + \underline{\underline{\tau}}_{\rm{CPA}}(\epsilon)\underline{\underline{\Delta}}_{a}(\epsilon)\right]^{-1}\underline{\underline{\tau}}_{\rm{CPA}}(\epsilon)\right)^{nn}.
    \label{eq:step1}
\end{equation}
Defining $\underline{\underline{\gamma}}_{a}(\epsilon) = \left[1 + \underline{\underline{\tau}}_{\rm{CPA}}(\epsilon)\underline{\underline{\Delta}}_{a}(\epsilon)\right]^{-1}$, it can be shown that
\begin{equation}
    \underline{\underline{\gamma}}_{a}(\epsilon)  = 1\;-\; \underline{\underline{\tau}}_{\rm{CPA}}(\epsilon)\underline{\underline{\Delta}}_{a}(\epsilon) \underline{\underline{\gamma}}_{a}(\epsilon) \label{eq:inversetrick}.
\end{equation}
Inserting (\ref{eq:inversetrick}), and using (\ref{eq:clusterimpurityeqn}), (\ref{eq:step1}) simplifies to
\begin{equation}
     \underline{\tau}^{nn}_{\rm{CPA}} = \underline{\tau}_{\rm{CPA}}^{nn}(\epsilon) - \sum_{a} c_a\left( \underline{\underline{\tau}}_{\rm{CPA}}(\epsilon)\underline{\underline{\Delta}}_{a}(\epsilon)\underline{\underline{\tau}}_a(\epsilon)\right)^{nn}.
\end{equation}
Hence,
\begin{equation}
    \sum_{a} c_a\left( \underline{\underline{\tau}}_{\rm{CPA}}(\epsilon)\underline{\underline{\Delta}}_{a}(\epsilon)\underline{\underline{\tau}}_a(\epsilon)\right)^{nn} = 0
    \label{eq:scf-condition}
\end{equation}
holds at self-consistency. However, for the initial guess and the first few iterations, this will be not be satisfied. Suppose at the $i^{th}$ and $(i + 1)^{th}$ iteration, the t-matrix for the medium is given by $\left[\underline{t}_{\rm{CPA}}(\epsilon)\right]^{i}$ and $\left[\underline{t}_{\rm{CPA}}(\epsilon)\right]^{i + 1}$ respectively. With significant algebraic manipulation, we re-phrase (\ref{eq:scf-condition}) as
\begin{widetext}
\begin{equation}
    \left[\underline{t}^{-1}_{\rm{CPA}}(\epsilon)\right]^{i + 1}  = \left[\underline{t}^{-1}_{\rm{CPA}}(\epsilon)\right]^{i}  + 
    \left[\underline{\tau}_{\rm{CPA}}^{nn}(\epsilon)\right]^{-1}\sum_{a} c_a\left( \underline{\underline{\tau}}_{\rm{CPA}}(\epsilon)\underline{\underline{\Delta}}_{a}(\epsilon)\underline{\underline{\tau}}_a(\epsilon)\right)^{nn}\left[\underline{\tau}_{\rm{CPA}}^{nn}(\epsilon)\right]^{-1}
    \label{eq:scheme}
\end{equation}
\end{widetext}
At self-consistency (\ref{eq:scheme}) reduces to (\ref{eq:scf-condition}). This iteration scheme produces an effective medium containing local behaviour. The scheme has an approximately linear form, which helps convergence. 

\subsection{Comparison with NL-CPA}
Other methods to include SRO in CPA were briefly discussed in the introduction of this paper. The most promising of these methods is the non-local NL-CPA. This section aims to highlight the difference between our cluster averaged CA-CPA and NL-CPA.

\paragraph{Effective Medium} 
While both NL-CPA and CA-CPA are self-consistent, NL-CPA also involves calculating a new structure constant which contains non-local behavior. This is not present in CA-CPA. In CA-CPA, there is a choice of calculating a self-consistent medium, or simply using the single-site CPA medium. The advantage of using the single-site CPA medium is simplicity and computational efficiency. 
\paragraph{Cluster Embedding}
In each NL-CPA iteration, the new effective medium is obtained by embedding and averaging over multiple real atom clusters. 
\begin{equation}
    \left<\underline{\tau}_{I}^{nm}(\epsilon)\right> = \underline{\tau}_{\rm{NL-CPA}}^{nm}(\epsilon)
\end{equation}
In contrast, CA-CPA only requires embedding a single cluster that is parametrized by the degree of SRO that we obtain from model or experiment. In self-consistent CA-CPA, only one cluster needs to be embedded per iteration. This is a significantly simpler approach as it is independent of the number of configurations and can be easily applied to multi-species high entropy alloys which may have a large configuration space.

\section{Results}
To demonstrate CA-CPA, we apply it to alloy systems with SRO at low temperatures. First we apply it to the Cu-Zn binary, a system whose SRO has been studied using both experimental and numerical techniques, including NL-CPA \cite{nlcpa-2,cuzn-expt,cuzn-expt-2,cuzn-dft}. We also apply it to AlCrTiV, to demonstrate that CA-CPA can easily deal with complex systems like high entropy alloys.
\subsection{Cu-Zn}
The CuZn BCC solid solution is a well-known example of a system that has an ordered BCC (B2) structure at low temperatures and transitions to a disordered BCC (A2) structure at approximately 750 K \cite{cuzn-expt,cuzn-expt-2}. This is a suitable system to test the CA-CPA because the inclusion of SRO will have a noticeable effect on the total energy and electronic density of states. 

\subsubsection{Energy Analysis}
\begin{figure}
\includegraphics{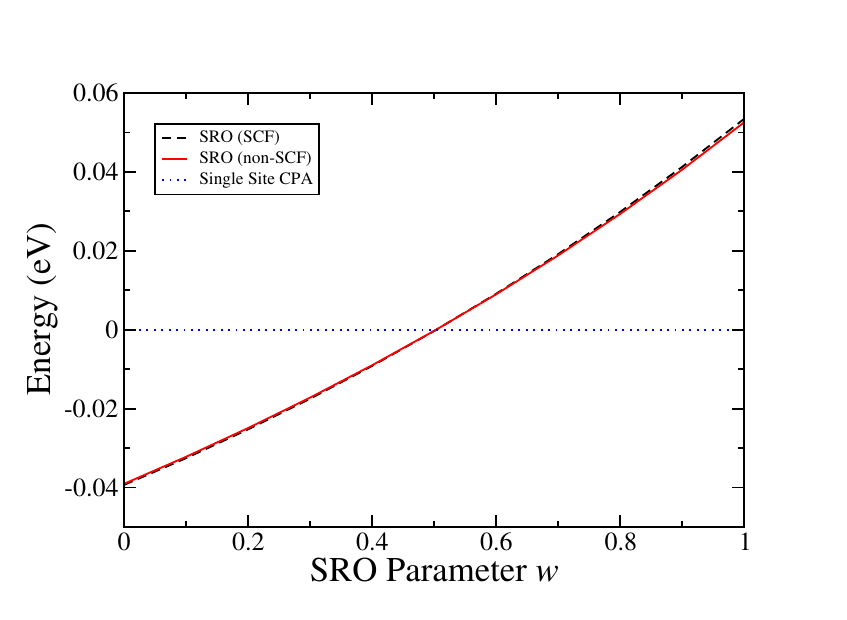} 
\caption{\label{fig:energyCuZn} Variation of the energy of CuZn with respect to the short range order parameter $w = w_{CuCu} = w_{ZnZn}$ (using single site CPA energy as the reference level), calculated using fully self-consistent and non-self-consistent CA-CPA. $w = 0$ corresponds to complete short range ordering up to nearest neighbors while $w = 1$ corresponds to short range clustering up to nearest neighbors}
\end{figure}

As a first test we compare the energy obtained from single-site CPA with the energy obtained from CA-CPA using different possible SRO parameters. Two different versions of CA-CPA have been used - a fully self-consistent CA-CPA (black dashed line) and the non-self-consistent version where the averaged cluster is embedded in single-site CPA medium (red line). Although there are four possible parameters for a binary, forcing the constraints $w_{ab} = w_{ba}$ and $\sum_{b} w_{ab} = 1$
results in only a single independent SRO parameter $w = w_{CuCu} = w_{ZnZn}$. Figure~\ref{fig:energyCuZn} shows the variation of energy with $w$. The energy is lowest when $w = 0$, i.e., unlike neighbors are preferred, corresponding to a B2 type ordering as expected. At $w = 0.5$, the ``averaged'' neighbor atoms are an equal mix of like and unlike atoms. This is similar to a completely disordered state, and this explains why the energy at $w = 0.5$ closely matches the single site CPA energy (denoted by the blue dotted line). Finally, as $w \xrightarrow{} 1$, corresponding to the unphysical case of short range clustering, the system becomes increasingly unstable. The difference between the self-consistent and non-self-consistent versions is of the order of 1 meV, which is an order of magnitude smaller than the energy change caused by SRO. Self-consistency does not make significant impact on the SRO calculation. Hence, the DOS calculation for CuZn has been done using the non-self-consistent CA-CPA in order to save computational time.

\subsubsection{Density of States}
\begin{figure}
\includegraphics{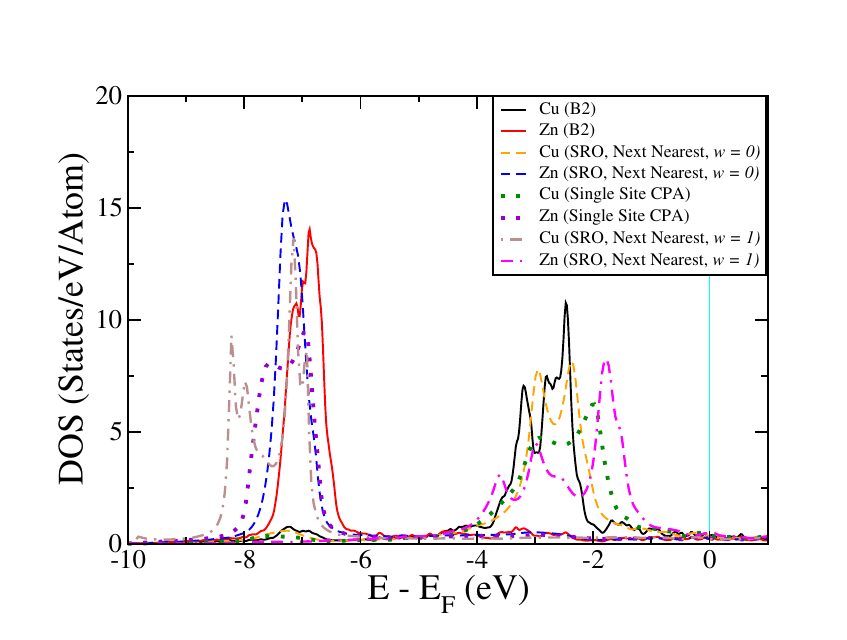} 
\caption{\label{fig:dosCuZn} Density of States for four different cases - the solid lines represent the pure B2 structure calculated using the KKR method, the dashed lines represent short range ordered ($w$ = 0 at nearest neighbor shell and $w = 1$ at next nearest neighbor) and short range clustered ($w$ = 1) up to next nearest neighbors using CA-CPA, and the dotted lines represent the A2 structure obtained from single site CPA.}
\end{figure}

As a second test we study the effect of SRO on the electronic density of states (DOS). As shown in Figure~\ref{fig:dosCuZn}, when $w=0$ the CA-CPA DOS better approximates the B2 DOS as compared to the single site CPA. The cluster embedding reduces the broadening of the single site CPA DOS. This is expected because the broadening is associated with disorder~\cite{disorder-dos}, and the CA-CPA reduces the disorder in the system. 

We can compare our results with the DOS calculated using NL-CPA \cite{nlcpa-2}. The main point of comparison is the size of the clusters embedded. Pair clusters were used in the NL-CPA study, and the effect of SRO was demonstrated by calculating the cluster component DOS. We have not done pair-cluster calculations, although it can be easily performed using CA-CPA. Instead, in our work, the DOS has been obtained by embedding a 15 atom cluster (central atom, 8 nearest and 6 next nearest neighbors).  Due to the larger cluster size, the difference in SRO DOS and single-site CPA DOS is more pronounced in our work.

\begin{figure}
\includegraphics{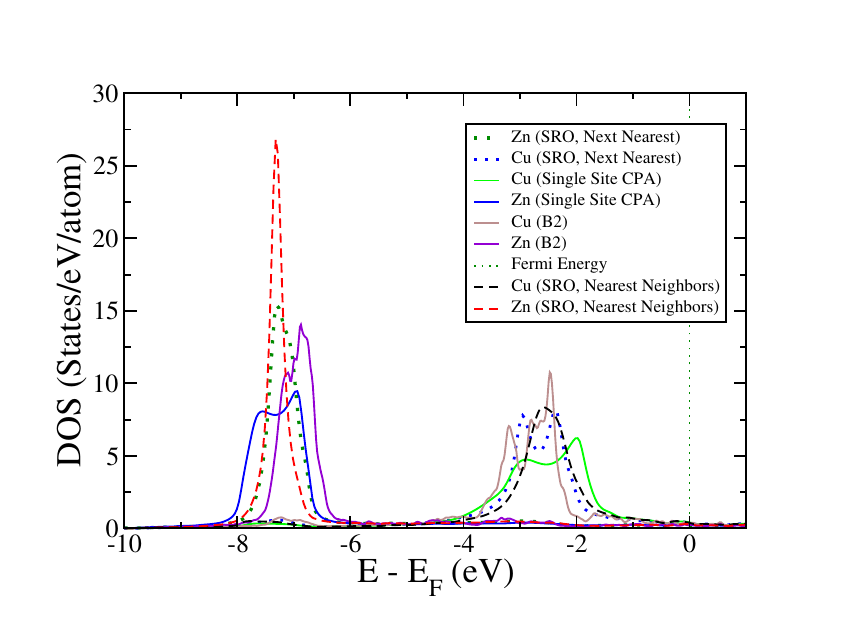} 
\caption{\label{fig:dosCuZn_compare} Comparing the DOS obtained using CA-CPA up to nearest ($w = 0$)  and next nearest neighbors ($w = 0$ for the near neighbor shell and $w = 1$ for the next nearest shell) against the pure B2 structure obtained using the KKR method, and the completely disordered structure obtained using single-site CPA.}
\end{figure}

We can also see the effect of varying the size of the averaged cluster in Figure \ref{fig:dosCuZn_compare}. The DOS obtained using CA-CPA up to next nearest neighbors is closer to the B2 DOS as compared to the DOS obtained using a nearest neighbor cluster. This seems to indicate that larger clusters reduce the randomness, and bring the system closer to an ordered structure.
\subsection{AlCrTiV}
The four-element AlCrTiV high entropy alloy displays B2 type ordering at low temperatures, with (Al, Cr) at one sublattice and (Ti, V) at the other~\cite{experimental_AlCrTiV}. As temperature rises and disorder increases, the energy gap between A2 and B2 structure reduces. We use this test case to show that CA-CPA scales well with the number of species present in the system and thus the number of possible configurations. For a 9 atom cluster (central atom plus 8 neighbors) in a 4 element random alloy, there are $4^9 = 262144$ configurations. Many of these configurations will not be unique. By considering space group symmetry, identical structures can be removed. The set of unique configurations will be smaller, but it still remains impractical to use any real atom cluster approach to study short range order in such a system. Using CA-CPA opens the door to modelling short range order in high entropy alloys. 

\subsubsection{Energy Analysis}
\begin{figure}
\includegraphics{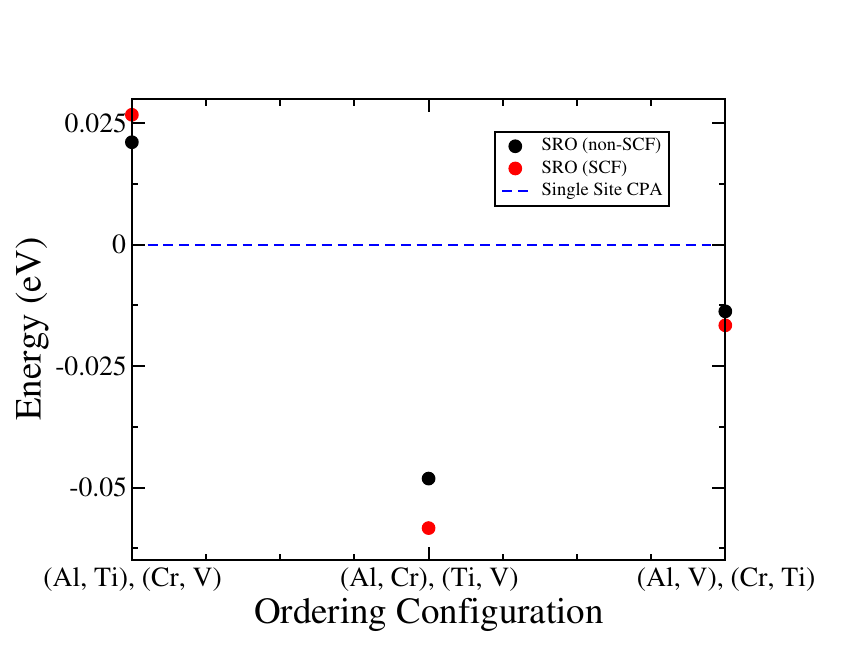} 
\caption{\label{fig:energyAlCrTiV} Energy of three possible B2 type orderings up to nearest neighbors (with single site CPA energy as the reference level)}
\end{figure}

For a 4-element high entropy alloy, there are 6 independent SRO parameters, and hence it is difficult to study how the energy varies with all of these 6 parameters. Instead we focus on three possible orderings -  \begin{enumerate*}[(a)] \item{AlCr-type ordering, where Al and Cr share the same sublattice (and of course, Ti and V share the other).} \item{AlV-type ordering, where Al and V share the same sublattice. } \item{AlTi-type ordering, where Al and Ti share the same sublattice.} \end{enumerate*}. Tables~\ref{Tab:AlCr}, \ref{Tab:AlV} and~\ref{Tab:AlTi} contain the SRO parameters used for the three orderings. The two CA-CPA modes have also been compared for all three orderings. The energies of these three structures are presented in Figure \ref{fig:energyAlCrTiV}. Here it can be seen that AlCr-type ordering has the lowest energy, which is the expected result \cite{experimental_AlCrTiV}. Self-consistency lowers the energy of AlCr and AlV-type ordering, and raises the energy of AlTi-type ordering. While the difference between self-consistent and non-self-consistent CA-CPA is larger for AlCrTiV as compared to CuZn, it is still small in comparison to the SRO energy change. As a result, the DOS calculation has been done using the non-self-consistent CA-CPA. 
\begin{table}[b]
\caption{SRO Parameters for B2 AlCr-type ordering}
\begin{ruledtabular}
\begin{tabular}{ccccc}
 $w$ & Al & Cr & Ti  & V \\
\hline
Al & 0 & 0 & 0.5 & 0.5 \\
Cr & 0 & 0 & 0.5 & 0.5 \\
Ti & 0.5 & 0.5 & 0 & 0 \\
V  & 0.5 & 0.5 & 0 & 0 \\
\end{tabular}
\end{ruledtabular}
\label{Tab:AlCr}
\end{table}
\begin{table}[b]
\caption{SRO Parameters for B2 AlV-type ordering}
\begin{ruledtabular}
\begin{tabular}{ccccc}
 $w$ & Al & Cr & Ti  & V \\
\hline
Al & 0 & 0.5 & 0.5 & 0 \\
Cr & 0.5 & 0 & 0 & 0.5 \\
Ti & 0.5 & 0 & 0 & 0.5 \\
V  & 0 & 0.5 & 0.5 & 0 \\
\end{tabular}
\end{ruledtabular}
\label{Tab:AlV}
\end{table}
\begin{table}[b]
\caption{SRO Parameters for B2 AlTi-type ordering}
\begin{ruledtabular}
\begin{tabular}{ccccc}
 $w$ & Al & Cr & Ti  & V \\
\hline
Al & 0 & 0.5 & 0 & 0.5 \\
Cr & 0.5 & 0 & 0.5 & 0 \\
Ti & 0 & 0.5 & 0 & 0.5 \\
V  & 0.5 & 0 & 0.5 & 0 \\
\end{tabular}
\end{ruledtabular}
\label{Tab:AlTi}
\end{table}

\subsubsection{Density of States}
\begin{figure}
\includegraphics{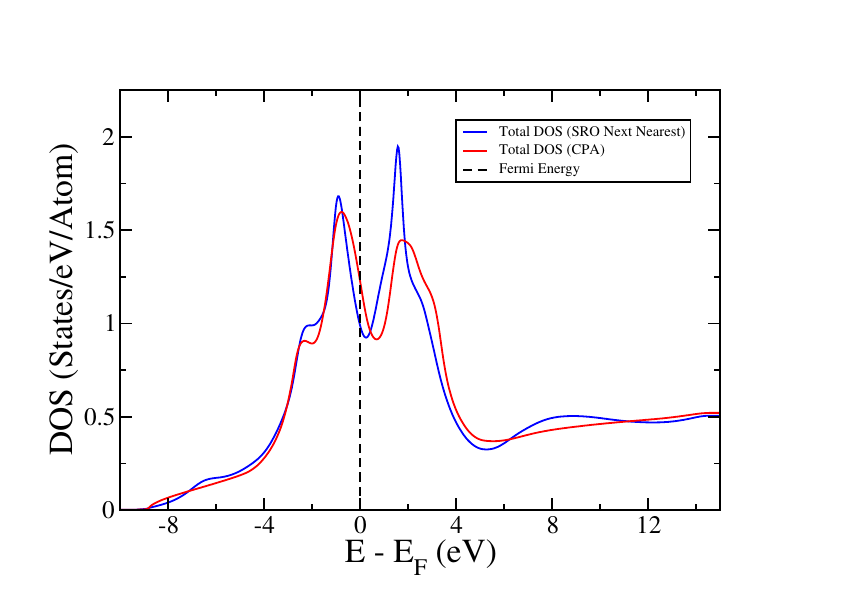} 
\caption{\label{fig:dosAlCr} Density of States for [(Al, Cr), (Ti, V)] ordered structure and disordered BCC structure.}
\end{figure}

The individual stability of the three configurations studied can also be discerned from the density of states
\paragraph{AlCr-type ordering} Figure \ref{fig:dosAlCr} shows the DOS for the disordered and the B2 ordered structure up to next nearest neighbors with Al and Cr on one sublattice. It can be seen that on the addition of SRO the DOS moves to the left, towards lower energies. Furthermore, the Fermi Energy has moved into the pseudogap. These observations indicate that this type of ordering has increased the stability of the system, which is the same conclusion derived from the energy analysis.

\begin{figure}
\includegraphics{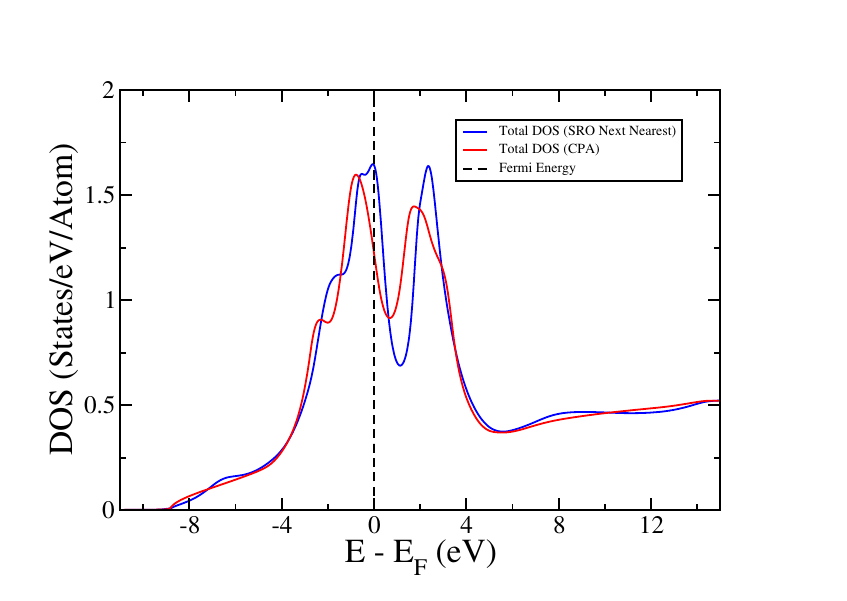} 
\caption{\label{fig:dosAlTi} Density of States for [(Al, Ti), (Cr, V)] ordered structure and disordered BCC structure.}
\end{figure}

\begin{figure}
\includegraphics{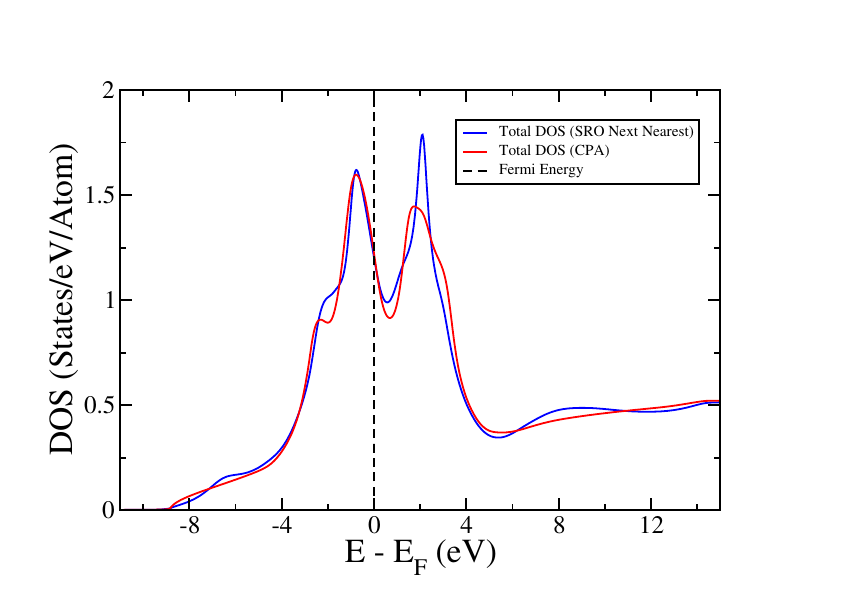}
\caption{\label{fig:dosAlV} Density of States for [(Al, V), (Cr, V)] ordered structure and disordered BCC structure.} 
\end{figure}

\paragraph{AlTi-type ordering} Figure \ref{fig:dosAlTi} shows the DOS for the disordered and the B2 ordered structure with Al and Ti on one sublattice. In this case, the opposite effect can be observed. The addition of SRO has shifted the DOS towards higher energies, and the Fermi Energy has moved away from the pseudogap. This type of ordering is unstable relative to the disordered structure.
\paragraph{AlV-type ordering} Figure \ref{fig:dosAlV} shows the DOS for the disordered and the B2 ordered structure with Al and V at one sublattice. In this case, there is no appreciable movement of the DOS on the addition of SRO and no clear conclusion on stability can be obtained from this plot. The implication is that this type of ordering is not as stable as the AlCr-type (clear shift towards lower energies) and not as unstable as AlTi-type ordering (clear shift towards higher energies), which matches the energy analysis plot.

\section{Conclusion}
We have introduced the Cluster Averaged Coherent Potential Approximation (CA-CPA) to include chemical short range order in the framework of KKR-CPA. This approach involves embedding a cluster consisting of ``averaged'' neighbor atoms constructed according to the SRO present in the system. The approach is independent of the number of configurations. It is simple to implement, and computationally efficient as only a single cluster per species needs to be embedded. The approach has been applied to a binary and a high entropy alloy, and the observations have been compared to existing results. A self-consistent extension to the CA-CPA has also been explored, and an iteration scheme to obtain a new effective medium has been derived. Results obtained using the self-consistent formalism have been compared with the non-self-consistent embedding. The code is part of the MuST package available on GitHub~\cite{MuST}.

The order parameters can be obtained in multiple ways. For binaries, there is only one independent SRO parameter. In such a case, it is not impractical to use a trial-and-error approach to find the SRO parameter that corresponds to the energy minimum. For more complex systems, this is less convenient, and experimental knowledge of SRO may be necessary to determine the order parameters. A third approach involves combining CA-CPA with a statistical physical method. Computer simulation~\cite{mcmd} can provide temperature-dependent predictions of short-range order parameters. The cluster variation method \cite{cvm} can combine order parameter-dependent energies from CA-CPA with a cluster expansion of the entropy to obtain the SRO parameters corresponding to the free energy minima across a temperature range. It would then be possible to calculate the transition temperature for systems that undergo an order-disorder transition.

\begin{acknowledgments}
This work is based on open-source ab initio software package MuST \cite{MuST}, a project supported in part by NSF Office of Advanced Cyberinfrastructure and the Division of Materials Research within the NSF Directorate of Mathematical and Physical Sciences under award number 1931367, 1931445, and 1931525. The conception and implementation of CA-CPA was supported by the Department of Energy under Grant No. DE-SC0014506. We acknowledge useful discussions with Nathaniel Hoffman in the early stages of code implementation.
\end{acknowledgments}

\end{document}